\documentclass[aps,prx,reprint,superscriptaddress]{revtex4-1}

\usepackage{array}
\usepackage{graphicx}
\usepackage{amsmath}

\usepackage{hyperref}
\usepackage{xprintlen}
\usepackage{enumitem}
\usepackage{color}
\usepackage{xcolor}
\usepackage{ulem}

\newcommand{\Fig}[1]{Figure~\ref{#1}}
\newcommand{\fig}[1]{Fig.~\ref{#1}}

\newcommand{\figs}[1]{Figs.~\ref{#1}}

\begin{document}

\title{Jamming and force distribution in growing epithelial tissue}

\author{Pranav Madhikar}
\affiliation{Department of Mathematics and Computer Science \&
  Institute for Complex Molecular Systems, Eindhoven University of
  Technology, 5600 MB, Eindhoven, Netherlands}
\author{Jan {\AA}str{\"o}m}
\affiliation{CSC Scientific Computing Ltd, K{\"a}gelstranden 14, 02150
  Esbo, Finland}

\author{Bj{\"o}rn
  Baumeier}
\affiliation{Department of Mathematics and Computer Science \&
  Institute for Complex Molecular Systems, Eindhoven University of
  Technology, 5600 MB, Eindhoven, Netherlands}

\author{Mikko Karttunen}
\affiliation{Department of Chemistry,  The University of Western Ontario, 1151 Richmond Street, London,
  Ontario, Canada N6A\,5B7
}
\affiliation{Department of Applied
  Mathematics,
  The University of Western Ontario, 1151 Richmond Street, London,
  Ontario, Canada  N6A\,5B7}
\affiliation{The Centre for Advanced Materials and Biomaterials Research,
  The University of Western Ontario, 1151 Richmond Street,
London, Ontario, Canada N6A\,3K7}

\date{\today}

\begin{abstract}
We investigate morphologies of proliferating cellular tissues using
a newly developed numerical simulation model for mechanical cell division
and migration in 2D. The model is applied to a bimodal mixture consisting of
stiff cells with a low growth potential and soft cells with a high growth potential;
cancer cells are typically considered to be softer than healthy cells.
In an even mixture, the soft cells develop into a tissue matrix and the
stiff cells into a dendrite-like network structure. When soft cells are placed inside
a tissue consisting of stiff cells (to model cancer growth), the soft cells develop to a fast growing tumor-like structure that gradually evacuates the stiff cell matrix.
The model also demonstrates 1) how soft cells orient themselves in the direction
of the largest effective stiffness as predicted by the theory of Bischofs and Schwarz
(Proc. Natl. Acad. Sci U.S.A., \textbf{100}, 9274--9279 (2003) and 2) that
the orientation and force generation continue a few cell rows behind the soft-stiff
interface. With increasing inter-cell friction, tumor growth slows down and cell death
occurs. The contact force distribution between cells is demonstrated to be highly
sensitive to cell type mixtures and cell-cell interactions, which indicates that
local mechanical forces can be useful as a regulator of tissue formation.
The results shed new light on established experimental data.
  
\end{abstract}


\maketitle

\section{Introduction}

Morphology and dynamics of proliferating cells are fundamental issues in cellular
development~\cite{Paluch2009,Li2012,Heisenberg2013,Ragkousi2014,Gibson2014,Sanchez-Gutierrez2015}. They are controlled by a number of factors, but from the physical point of view, morphology is tightly coupled to inter-cellular force transmission, see e.g. Refs.~\cite{Rauzi2008,Tambe2011,Sanchez-Gutierrez2015}.
Mechanical forces have been shown to be important in tissue healing after damage~\cite{Trepat2009} and
cancer development, and it has been suggested that tumor growth may even become arrested by inter-cellular mechanical forces~\cite{Helmlinger1997,Kumar2009,Alessandri2013}.
In addition to uniform structures, a plethora of structures with various mechanisms and division modes have been suggested but the issue remains largely
unresolved~\cite{Pilot2005,Gibson2006,Guillot2013,Herrero2016}.

Among the many complications in investigating force transmission are that at
their embryonic state, cells may not yet have developed junctions and
may display more fluid-like behavior, and that cell-cell adhesion
depends on the cell type and type of adhesion (focal or non-focal)~\cite{Lecuit2005,
  Rauzi2008, Guillot2013,Helvert2017}. Junctions are crucial in
cell-to-cell stress transmission~\cite{Rauzi2008, Liu2010, Tambe2011,Pontani2012} but it is, however,
challenging to probe the individual junctions experimentally.
In addition, the different cellular level
signaling mechanism can be coupled depending on local properties and environment, that is, chemical signaling
based on molecular interactions
and mechanotransduction may depend on each other, see e.g. Ref.~\cite{Zhang2012f}.

From a coarse-grained point of view, i.e., ignoring chemical details and treating cells as elastic objects, cellular systems can be seen as
disperse grand canonical soft colloidal systems under evolving pressure.
Several studies have tried to capture aspects
of growing soft matter systems~\cite{Astrom2006,Jones2012,Gonzalez-Rodriguez2012, Kursawe2018} but even in simple
systems many fundamental questions remain open including the precise nature
of colloidal phase diagrams when colloids are soft with size
dispersity~\cite{Fernandez2007}, and structure selection via
self-assembly~\cite{Frenkel2011}. Cellular systems are even more
complex since they exhibit additional behaviors such as
growth and division, have varying mechanical properties (e.g. elasticity
and cell-cell adhesion) and their responses to external stimuli may be sensitive to the local environment.

\begin{figure*}[ht]
  \centering
  \includegraphics[width=0.85\textwidth]{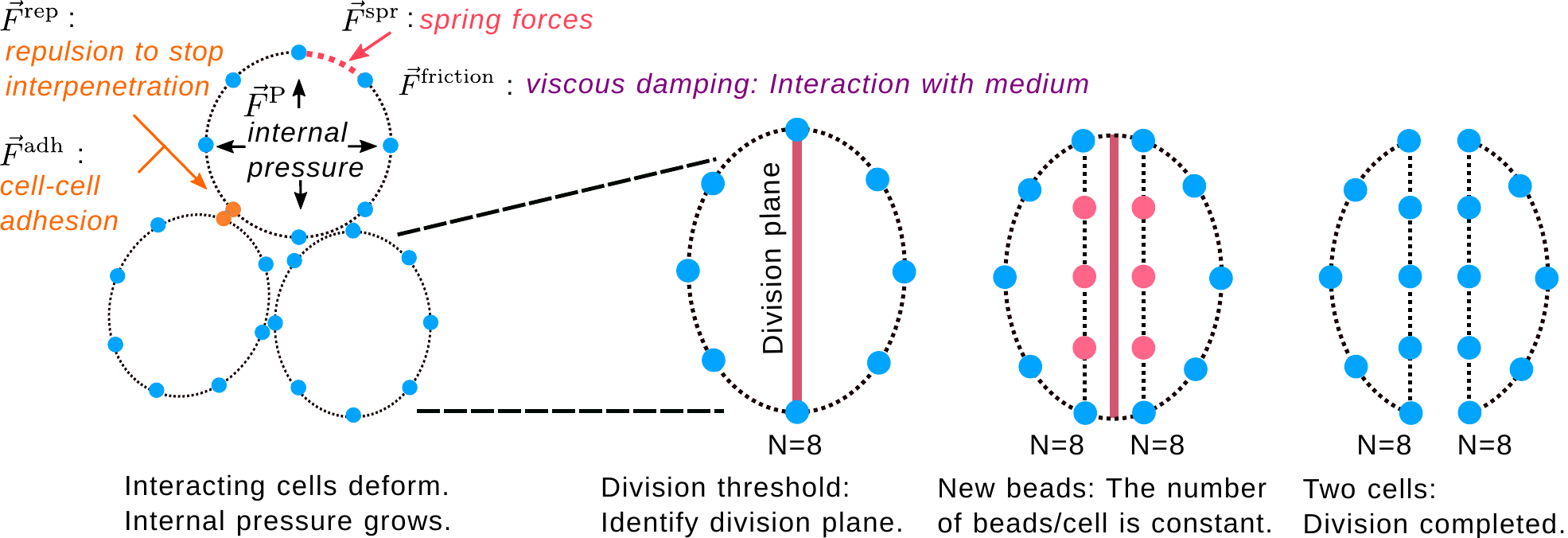}
\caption{Illustration of the model and the forces (Eq.~\ref{eq:force_field}). As the cells grow and deform, they reach the division threshold. At that time, a division plane is identified based on the long-axis (Hertwig's) or some other selected rule. As the cell divides, new beads must be added to preserve the topology of the cells. Once division is completed, two child-cells are produced. Colors indicate the different forces and the new beads that need to be created at the time of division.}
  \label{fig:cellmodel}
\end{figure*}

Dimensionality has an important role in regulation of intra- and
inter-cellular forces at different levels, see e.g.
Refs.~\cite{Charnley2013,Hernandez-Hernandez2014,Lesman2014,Osterfield2017,Owen2017}.
Systems such as epithelial tissues and \textit{Dro\-sophila} wing discs,
are inherently two dimensional which gives them distinct morphological
properties due to the nature of cellular packing, and transmission of and response to
forces~\cite{Gibson2009,Guillot2013,Gibson2014}.
In addition, jamming can be very strong in two dimensions;
understanding the effects of stiffness, density and inter-membrane friction is
crucial for being able to determine how jamming emerges in cellular systems~\cite{Sadati2013,Reichhardt2014}.
Besides being important in understanding the mechanisms of cell movement
under pressure~\cite{Tambe2011}, such situations have been proposed to be important in tumor
growth~\cite{Helmlinger1997,Kumar2009} -- cancer cells are often softer than healthy
cells~\cite{Alibert2017} although the opposite has also been reported~\cite{Suresh2007}.
Our main focus is on the above effects in systems consisting of hard cells in a soft matrix and vice versa.
We use our previously introduced two dimensional computational model
called \textit{EpiCell2D} (Epithelial Cell 2D)~\cite{Mkrtchyan2014}.
The model is summarized below in Models and Methods and described in full detail in Ref.~\cite{Mkrtchyan2014}.
Cell stiffness, its measurements and connection to cancer metastasis have been recently reviewed by Luo \textit{et al.}~\cite{Luo2016}.

Two dimensional cellular systems have been studied with a
number of computational methods including off-lattice vertex
models~\cite{Honda2004, Farhadifar2007, Hufnagel2007, Fletcher2014,
  Sussman2018, Staple2010, Fletcher2010, Gibson2006, Nagpal2008,
  Patel2009,Alt2017}, and Voronoi tessellation or Delaunay triangulation
based models~\cite{Schaller2004, Meyer-Hermann2005,
  Meyer-Hermann2008, Beyer2009, Radszuweit2009}. Such models
approximate cell membranes as edges or planes and
they do not include inter-cellular forces.
The immersed boundary method
of Rejniak \textit{et al.}~\cite{Rejniak2007} which explicitly models
inter-membrane interactions is better suited to problems that require
inter-cellular forces.
Other methods include lattice-based methods  such as the Cellular
Potts Model~\cite{Graner1992, Szabo2013,
  Shirinifard2009, Merks2005}. Although versatile, they describe
cellular interactions with scalar energy terms, making it impossible
to study forces between cells unless they are amended.
Phase-field modeling, based on defining
a free energy functional and an order parameter (or several parameters) and
is a relatively new approach to cell division and migration~\cite{Nonomura2012,Palmieri2015,Palmieri2019,Jiang2019}.

The models that describe cells as discrete entities with individual properties can be further
classified as either off-lattice~\cite{Liedekerke2018} or lattice-based~\cite{Drasdo2018} agent models.
For example, the former include the models of Rejniak \textit{et al.}~\cite{Rejniak2007, Rejniak2008}
and our current model, and the latter cellular automata and Potts models.
In off-lattice models cells are typically deformable and forces or energies
are used in determining their behavior whereas in lattice-based models updates are done based on pre-determined rules rather than forces and the cellular shapes or topologies are typically
fixed. Both approaches have their advantages and caveats. A comprehensive review of agent-based cellular models is provided by Van Liedekerke \textit{et al.}~\cite{Liedekerke2015} and
a review of mechanobiological aspects and morphology during the whole cell cycle is provided by Clark and Paluch~\cite{Clark2011}.

\section{Model and methods}

We employ the two-dimensional \textit{EpiCell2D} model to study tissue morphologies.
Full  details, derivation, and parameter mapping with all parameters (force field) are
provided in Ref.~\cite{Mkrtchyan2014} and not repeated here. Below, we summarize the salient
features and provide the new additions.
In the context of the models discussed above, \textit{EpiCell2D}
can be classified as an off-lattice agent-based model; cells
are well-defined entities, have individual properties, they can deform, divide
and forces are used to update their positions and motions.

In \textit{EpiCell2D} the cell membrane is discretized as beads connected by elastic bonds
of stiffness $K^\mathrm{spr}_i$ to form a closed loop, see Fig.~\ref{fig:cellmodel}.
In our simulations, the cells had 76 beads (Fig.~\ref{fig:cellmodel} has only eight
for clarity). The model has been previously tested~\cite{Mkrtchyan2014} using
10-100 beads per cell and it was found that the number of beads has no effect
on results if $N\ge40$. The model parameters (details in Ref.~\cite{Mkrtchyan2014})
were mapped using a cell diameter of 10\,$\mu$m, mass of $10^{-12}$\,kg~\cite{Phillips2012}
and Young's modulus from mitotic HeLa cells~\cite{Stewart2011}. The orientation
of the division plane was chosen randomly such that it passed through the center
of mass of the cell.

In  \textit{EpiCell2D}, the force field is defined as a sum of intra-cellular,
inter-cellular, and cell-medium interactions, see Fig.~\ref{fig:cellmodel}.
The intra-cellular terms include the internal pressure ($\vec{F}^\mathrm{P}$)
and spring forces ($\vec{F}^\mathrm{spr}$) which provide a mean-field description
for the components that give cells their integrity (cell cortex contractility),
that is, the cell membrane and cytoskeleton. The inter-cellular terms define
the interactions between the cells. The first of them ($\vec{F}^\mathrm{rep}$)
is repulsion to prevent cells from penetrating each other and the second
term ($\vec{F}^\mathrm{adh}$) describes cell-cell adhesion. In real cells
adhesion occurs, e.g., due to lipids and different adhesive proteins
depending on the cell type. For a recent discussion on the physical aspects
of adhesion, see, e.g., Schulter~\textit{et al.}~\cite{Schluter2014} and
van Helvert \textit{et al.}~\cite{Helvert2017}. The final term is
medium-cell interaction (friction). With these, the force field can be
given in the following general form
\begin{equation}
  \label{eq:force_field}
  \vec{F} = m\ddot{\vec{r}} =
  \underbrace{\vec{F}^\mathrm{P}
  + \underbrace{\vec{F}^\mathrm{spr}}_\mathrm{bonded}}_\mathrm{intra-cellular}
  +  \underbrace{\vec{F}^\mathrm{rep} + \vec{F}^\mathrm{adh}}_\mathrm{cell-cell}
  +  \underbrace{\vec{F}^\mathrm{friction}}_\mathrm{cell-medium} .
\end{equation}
The functional forms of the terms and all the parameters (14 in total)
are given in Mkrtchyan \textit{et al.}~\cite{Mkrtchyan2014} and are not repeated here.


\begin{figure*}[ht]
  \centering
  \begin{minipage}[b]{0.3\textwidth}
    \includegraphics[width=51mm]{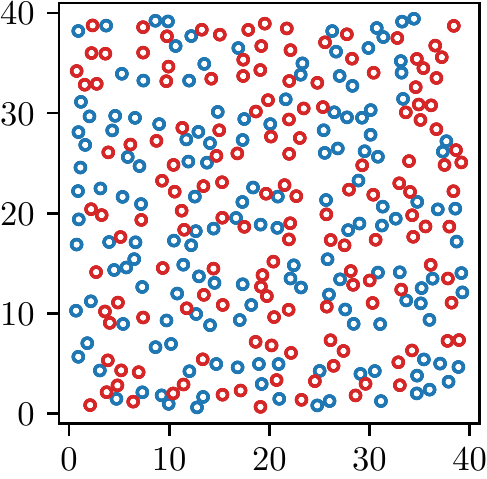}
        \put(-125,129){\colorbox{white!100}{\textbf{(a)}}}
  \end{minipage}%
  \begin{minipage}[b]{0.3\textwidth}
    \includegraphics[width=5.1cm]{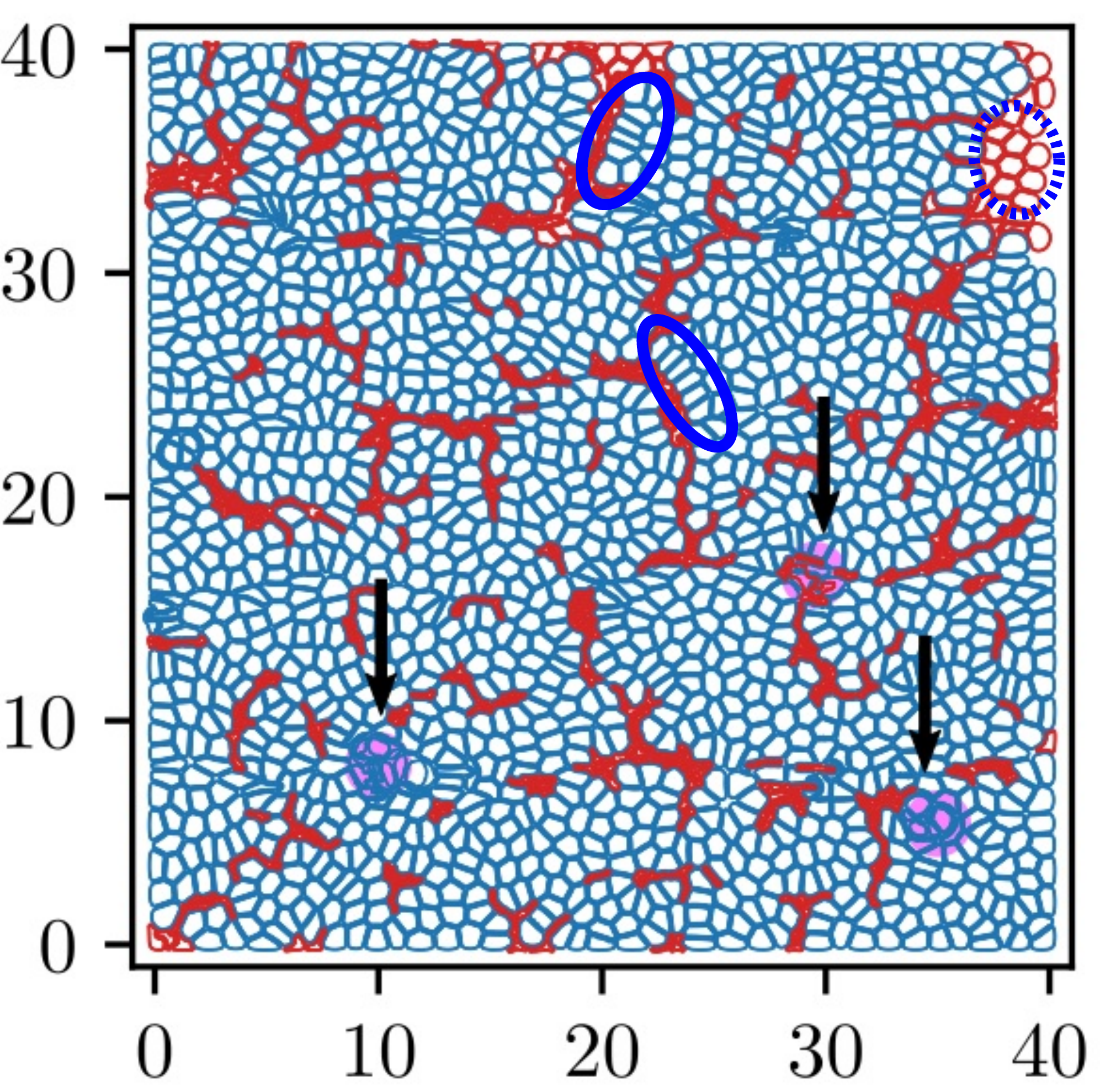}
        \put(-126,129){\colorbox{white!100}{\textbf{(b)}}}
  \end{minipage}%
  \begin{minipage}[b]{0.3\textwidth}
    \includegraphics{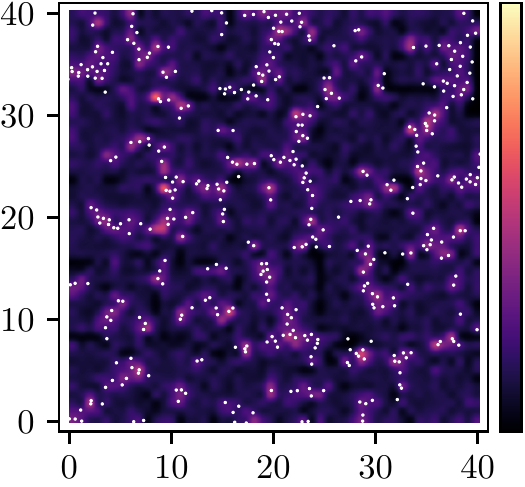}
    \put(-132,129){\colorbox{white!100}{\textbf{(c)}}}
    \put(2,15){0\,nN}
    \put(2,132){150\,nN}
  \end{minipage}
  \caption{
    Influence of stiffness differences in mixtures of stiff (red) and soft (blue) cells.
    The X- and Y- axes represent spatial coordinates in 2D in
    units of 10\,$\mu$m. (a) Initial configuration. Growth is simulated
    from this state onward until confluence
    with inter-membrane
    friction coefficient $\mu = 0.0\, \mu$g/s.
    (b) A confluent tissue of
    soft and stiff cells. Stiff cells form dendrite or vein-like
    structures in a matrix of soft cells. The regions marked
    with light purple and arrows are areas where cells interpenetrate and
    cell death occurs.
    The areas indicated with blue ovals show how
    the soft cells orient themselves at the boundary of stiff cells.
    The area surrounded by an oval with dashed lines show how
    the stiff cells retain their shapes when away from boundary.
    These issues are elaborated in Discussion.
    (c) Contact force distribution in the same tissue.
    Large contact forces
    are located at stiff cells and at boundaries between soft
    and stiff cells. White markers are the centers of masses of
    the stiff cells.}
  \label{fig:force_distrib}
\end{figure*}

In  \textit{EpiCell2D} cellular growth is controlled by a growth
pressure. This is motivated by the fact that cells have mechanisms to control
their internal hydrostatic pressure particularly before division~\cite{Stewart2011}.
Division is triggered by a threshold in cell area (above which cells divide).
We would also like to note that different criteria can be used. As pointed out but Streichan
\textit{et al.}~\cite{Streichan2014} at least area and growth rate are possible
criteria for triggering cell division.

Importantly, \textit{EpiCell2D}  allows the \textit{aggregate topology}
(the polygonal distribution) to vary spontaneously~\cite{Mkrtchyan2014}; note
that \textit{cellular topology} is fixed, that is, the number of nodes per
cell must remain constant upon division, Fig.~\ref{fig:cellmodel}.

One important issue in modeling cells is cell death. 
The role of cell death in tissue dynamics has been discussed in
numerous publications, a recent perspective is given by Green~\cite{Green2019}.
Different approaches have been taken to include cell death in computational models. In some models
that is done as a probability for a cell to disappear~\cite{Podewitz2015} or
via a tunable cell cycle rate~\cite{Czajkowski2019}. In the current study,
cell death occurs due to local stresses. All of these approaches can be
justified as cell death is a very complex and multifaceted phenomenon. Our choice
is based the observations of Chen \textit{et al.}~\cite{Chen1997} and Streichan
\textit{et al.}~\cite{Streichan2014} that mechanical constraints, in particular cell shape
and local stress, are critical factors and determinants for cell cycle and death.
Similar mechanisms have also been proposed and analyzed by Shraiman~\cite{Shraiman2005}. 

Previously, we focused on model development, parameter mapping and verification
against experiments using only one type of cells~\cite{Mkrtchyan2014}.
Due to the large parameter space (14 in total), it is not feasible, however,
to attempt to map a phase diagram. As a new direction, we extend
\textit{EpiCell2D} for simulations of different cell types using two
simple approaches with parameters based on experiments:
\begin{enumerate}
  [noitemsep]
\item changing cell stiffness, and
\item changing the cell-cell friction
  between cell membranes; in \textit{EpiCell2D} cell membrane and
  cytoskeleton are treated as a coarse-grained single object.
\end{enumerate}

Modification 1) allows for simulations of different cell types.
As mentioned above, cancer cells are typically softer than the
matrix cells and softness, or higher malleability, is typically
associated with the invasiveness of cancer cells~\cite{Luo2016,Alibert2017}.
This has recently been challenged by Nguyen~\cite{Nguyen2016a}
\textit{et al.} who measured Young's modulus of pancreatic cancer
cells using different cell lines and found the stiffer (than the
matrix cells) cells to be more invasive than the softer cancer cells.
Whether this is purely mechanical or due to simultaneously occurring
biological processes remains unclear; simulations using 
\textit{EpiCell2D}
indicate that stiffer cells migrate easier following in the wake
of a leader while softer cells collectively evacuate stiffer cells
due to aggressive growth.
This is in excellent agreement with the findings of Trepat
\textit{et al.} who showed that collective effects
are essential cell migration~\cite{Trepat2009}.
We will return to this issue in Results as well as in Discussion.

Here, we use two types of cells:  \textit{(i)} \textit{Stiff} cells
with a low growth potential with stiffness
$K^\mathrm{spr}_1 \!= \! 4$~$\mu\mathrm{N}/\mu\mathrm{m}$.
The low growth potential means that the cell membrane is so stiff
that the applied pressure is barely enough to grow the cell to
a size above the division threshold.
\textit{(ii)} \textit{Soft} cells with stiffness
$K^\mathrm{spr}_2\!=\! 1$~$\mu\mathrm{N}/\mu\mathrm{m}$.
These cells have a high growth potential which means that cell
membrane stiffness is low and the cell area can easily grow
beyond the division size; due to the lower elastic modulus,
the growth rate of a soft cell is higher even if the internal
pressure is the same as for a stiff cell. Note also that in
\textit{EpiCell2D} the plasma membrane and cytoskeleton
are treated as a coarse-grained single object referred
to as the membrane.

Modification 2) allows for comparisons of systems of cells
with different inter-membrane friction coefficients
(term $\vec{F}^\mathrm{adh}$ in Eq.~\ref{eq:force_field}
and Fig.~\ref{fig:cellmodel}). The importance of cell-cell
friction in mechanotransduction has recently been reviewed by
Angelini \textit{et al.}~\cite{Angelini2012}.

In Eq.~\ref{eq:force_field}  inter-membrane friction is
modeled via term
$
-\mu\vec{v}_{ij},
$
where $\mu$ is the friction coefficient and $\vec{v}_{ij}$ is
the relative velocity between two membranes
tangential to the cell that the bead $i$ belongs to.
We compare systems with $\mu\!=\!0.0$~$\mu$g/s, that is,  cells
do not interact very much with their neighbors, and strongly
interacting cells with $\mu\!=\!200.0$~$\mu$g/s.
In the simulations,
both open and closed boundaries were used.

\section{Results}

\begin{figure*}[ht]
  \centering
  \begin{minipage}{0.30\textwidth}
    \includegraphics{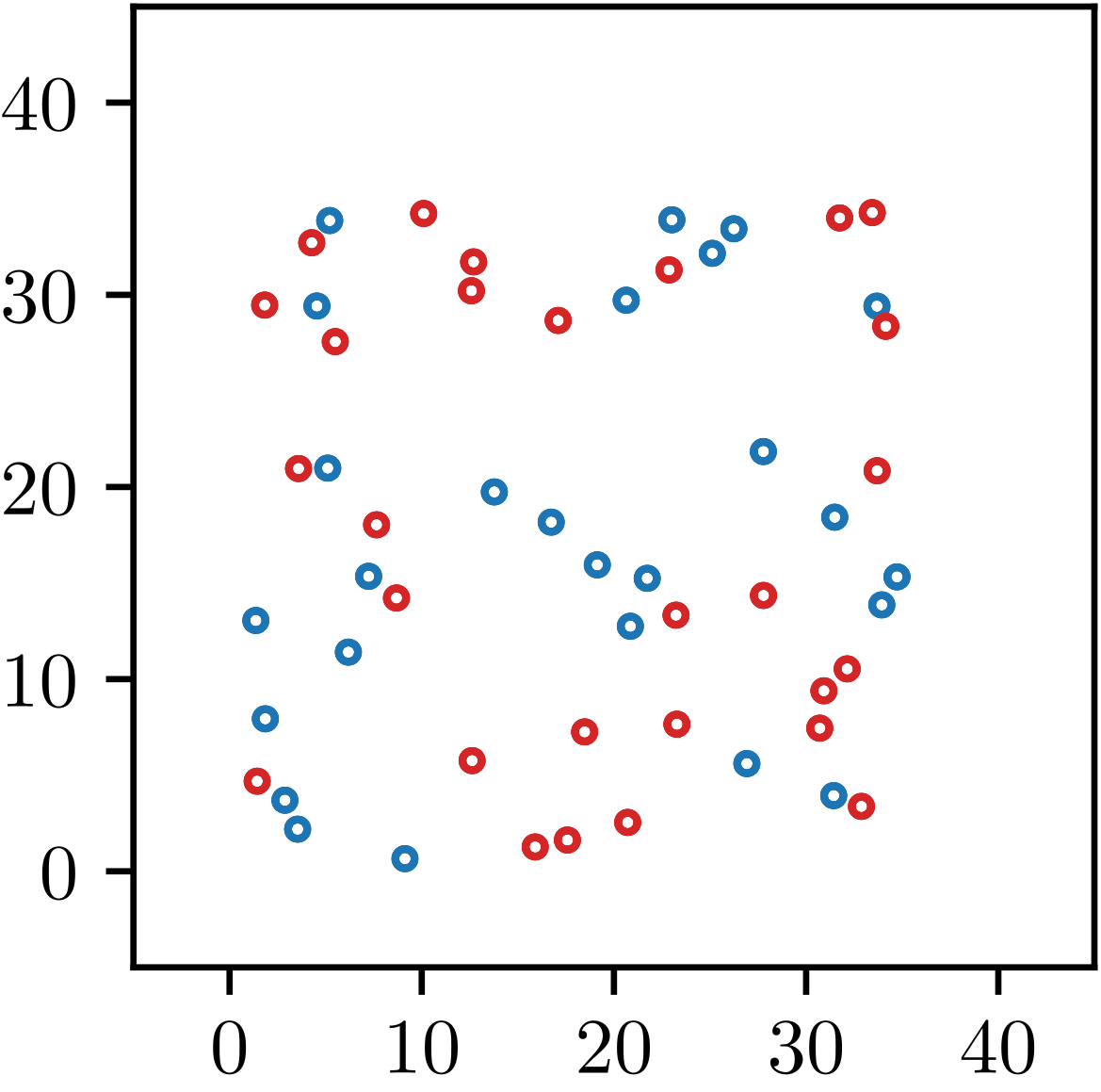}
    \put(-120,125){\textbf{(a)}}
  \end{minipage}%
  \begin{minipage}{0.30\textwidth}
    \includegraphics{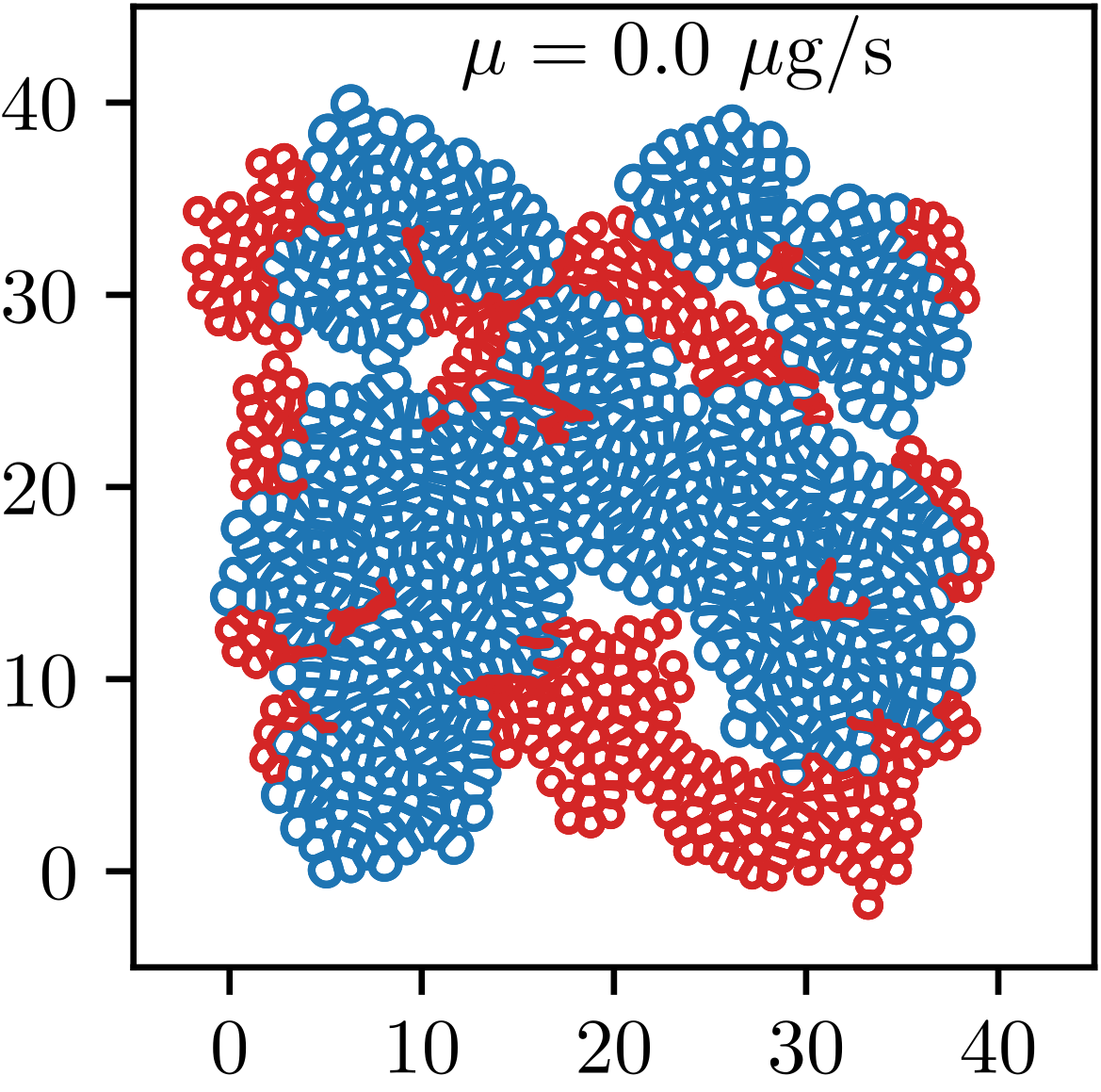}
        \put(-120,125){\textbf{(b)}}
  \end{minipage}%
  \begin{minipage}{0.30\textwidth}
    \includegraphics{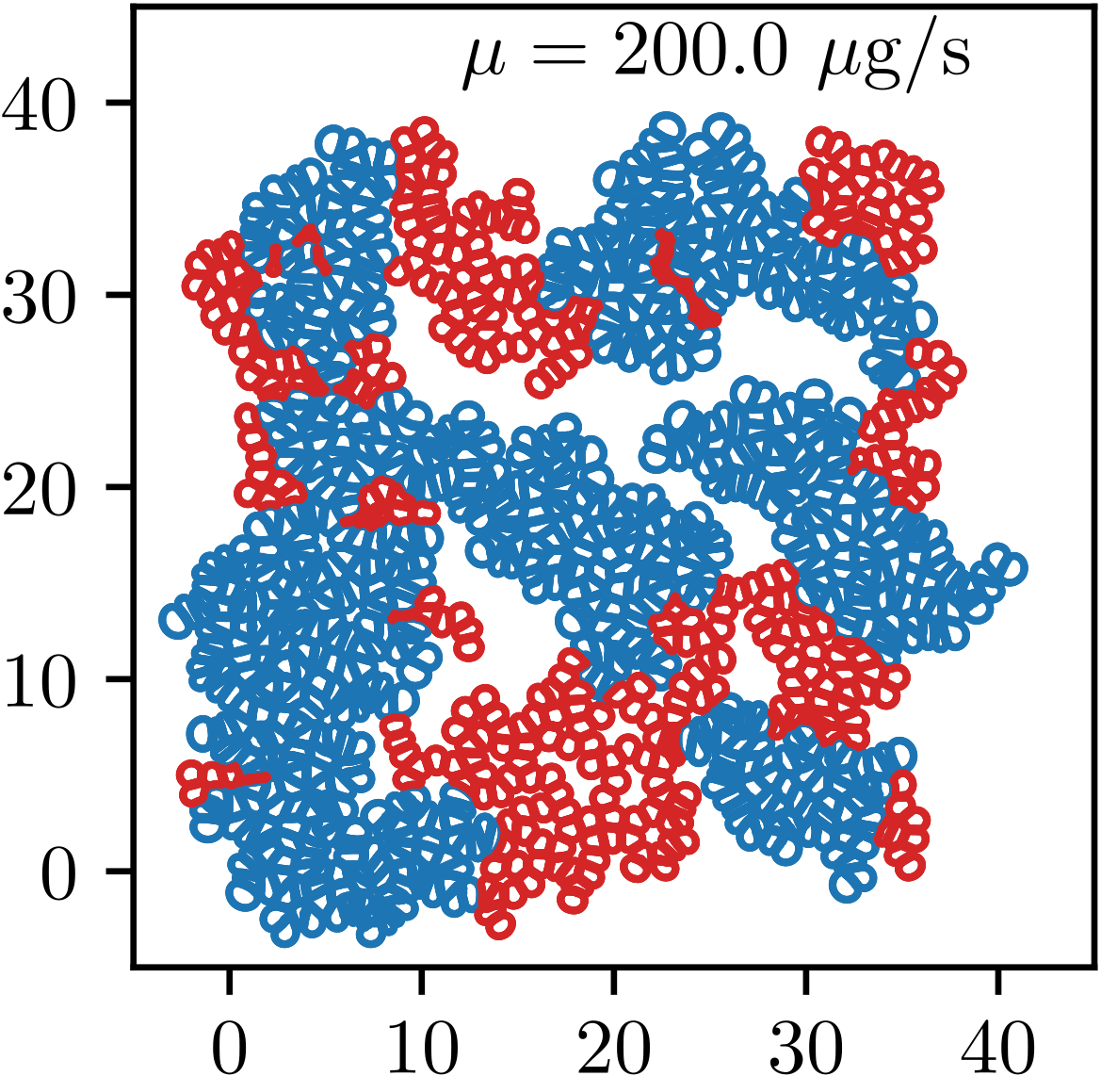}
        \put(-120,125){\textbf{(c)}}
  \end{minipage}
  \caption{Influence of inter-membrane friction on simulated cell morphologies. The X- and Y-axes represent spatial coordinates in 2D in
    units of 10 $\mu$m. (a) Initial conditions used in both (b) and (c).
    Stiff cells are shown in red and soft ones in blue. The two are present
    in equal proportions.
    (b) and (c): Morphologies after 10 division cycles in the cases of zero
      inter-cellular friction (b) and high friction ($\mu=200.0$~$\mu$g/s) (c).
      Growth is faster at $\mu=0.0$~$\mu$g/s as indicated by the larger number
      of cells after the same number of division cycles. Morphologically,
      the high inter-membrane friction system
      is more porous. The friction-less system corresponds to very
      early stages of development when junctions have not yet developed.
    }
  \label{fig:sim_hl_mu}
\end{figure*}

We focus on systems with two cell types, \textit{Stiff} and \textit{Soft},
present simultaneously. \textit{Soft} cells represent tumor cells based
on the fact that cancer cells tend to be softer~\cite{Alibert2017},
  Young's modulus for cancer cells is typically $\approx \!0.5$\,kPa and whereas
  for normal cells it often $\approx \!\!1.0\!-\!2.0$\,kPa (although variations are large,
  see, e.g., Ref.~\cite{Helvert2017}).
In real cancer cells this effect can be enhanced by a lesser number (or lack) of
adhesion proteins. However, although cancer cells are generally
assumed to grow faster than healthy cells, measurements are not
trivial as different metrics can be used~\cite{Mehrara2007}.
An additional complication is the fact that in healthy tissue
growth is regulated whereas cancer cells typically lack such regulation.
For a review of properties of cancer cells, see, e.g., Hanahan
and Weinberg~\cite{Hanahan2011}
and for biomechanical aspects
Fritsch \textit{et al.}~\cite{Fritsch2010}.

The initial setups were created by placing equal proportions of
\textit{Stiff} (red) and \textit{Soft} (blue) cells randomly,
see \Fig{fig:force_distrib}a; the simulations were repeated several
times and the results did not depend on the initial conditions.
\Fig{fig:force_distrib}b shows the tissue structure at the end
of one of the simulations. The soft cancer cells have invaded
the space while the stiff cells (red) have been compressed into
dendrite-like structures. Another distinct feature is that the
cells interpenetrate in the regions  marked with light purple
and arrows in \fig{fig:force_distrib}b. Although this behavior
may seem as an artifact, it occurs in diverse systems as has
been shown by Eisenhoffer \textit{et al.} for canine, human and
zebrafish epithelial cells~\cite{Eisenhoffer2012} and discussed
at length by Guillot and Lecuit~\cite{Guillot2013} (see Fig.~2
in Ref.~\cite{Guillot2013}). \Fig{fig:force_distrib}c shows the
average contact forces between cells.  The white dots show
the centers of masses of the stiff cells. The contact force peaks
correlate highly with the locations of the stiff cells indicating
that the \textit{Soft} cells overwhelm the \textit{Stiff} cells as the
tissue grows.

These interpenetrating cells
is where cell death occurs. Although the dead cells are not physically removed (to keep
the simulation code faster), they collapse and occupy only minimal space as defined by
the non-overlap of the surface nodes.

Next, we examine if the collapse of stiff cells can be mitigated
by making their interactions stronger. This is  done by
changing the magnitude of inter-membrane friction $\mu$. Since cells
need space to grow, they need to slide past each other into
empty regions. Higher friction, however, induces jamming and thus
reaching the division threshold takes a much longer time. The softer
cells will also need to counteract this effect to be able to grow.

\begin{figure}[ht]
  \centering
  \begin{minipage}{\columnwidth}
  \centering
  \includegraphics[width=0.95\columnwidth]{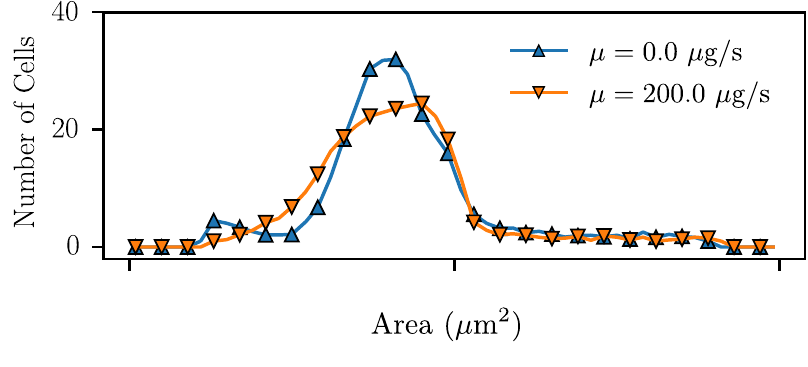}
    \put(-200,90)
    {\textbf{(a)}}
  \end{minipage}\\
  \begin{minipage}{\columnwidth}
    \centering
    \hspace{6pt}
    \includegraphics[width=0.92\columnwidth]
    {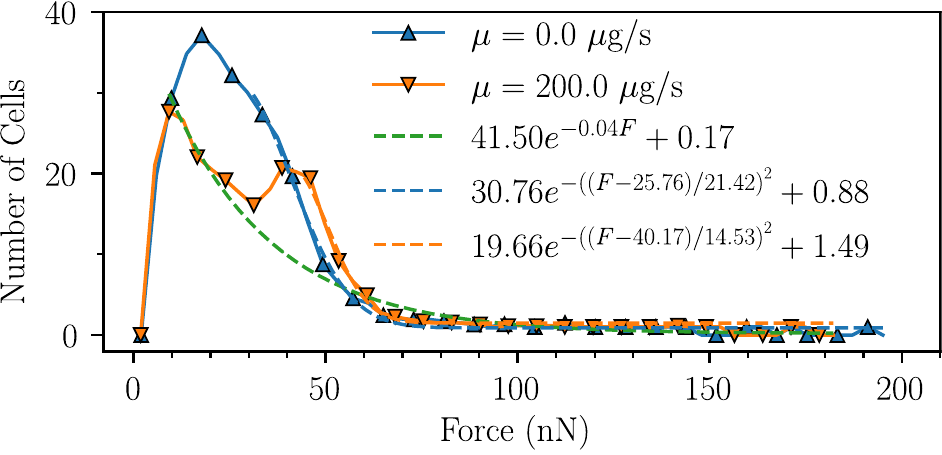}
    \put(-200,94)
      {\textbf{(b)}}
  \end{minipage}
  \caption{
    (a) Area and (b) cell-cell contact force  (denoted by $F$) distributions
      at low ($\mu\!=\!0.0$~$\mu$g/s)
      and high ($\mu\!=\!200.0$~$\mu$g/s) inter-membrane
      friction, corresponding to Figs.~\ref{fig:sim_hl_mu}b and
      c, respectively. (a)
    The peak at low areas  ($A\approx 20$\,$\mu$m$^2$) corresponds to collapsed
    stiff cells. The dashed lines in (b) are fits with the green line being
    a fit to the $\mu=200.0$~$\mu$g/s case ignoring
    the second peak. Solid lines are guide to the eye.
}
  \label{fig:area_force_hist}
\end{figure}

\begin{figure}[h!t]
  \centering
  \begin{minipage}{0.5\columnwidth}
    \includegraphics{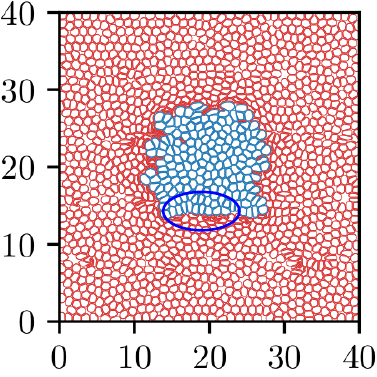}    
    \put(-91,112) {\textbf{(a)}{\bf \color{black}{$\mathbf{\boldsymbol{\mu}\!=\!0.0}$~$\boldsymbol{\mu}$g/s}}}

  \end{minipage}%
  \begin{minipage}{0.5\columnwidth}
    \includegraphics{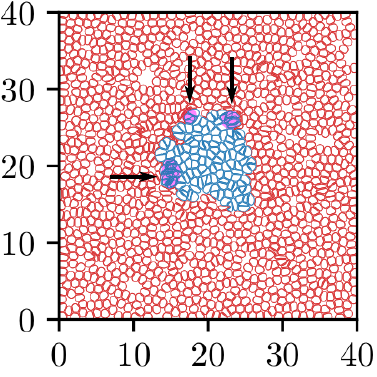}
    \put(-91,112) {\textbf{(b)}{\bf \color{black}{$\mathbf{\boldsymbol{\mu}\!=\!200.0}$~$\boldsymbol{\mu}$g/s}}}
      \end{minipage}
  \caption{
    Configurations  at confluence for the case of
    inclusion of \textit{Soft} (blue)
    cells in a matrix of \textit{Stiff} (red) cells when
    a) $\mu=0.0$~$\mu$g/s and b) $\mu=200.0$~$\mu$g/s.
    The areas indicated with blue ovals show how
    the soft cells orient themselves at the boundary of stiff cells.
     Purple regions and arrows: areas were cells interpenetrate and death could
    occur.
    X and Y axes: spatial coordinates  in units of 10\,$\mu$m.
  }
  \label{fig:snap_area_force}
\end{figure}

\begin{figure}[h!t]
  \centering
  \includegraphics[width=1.0\columnwidth]{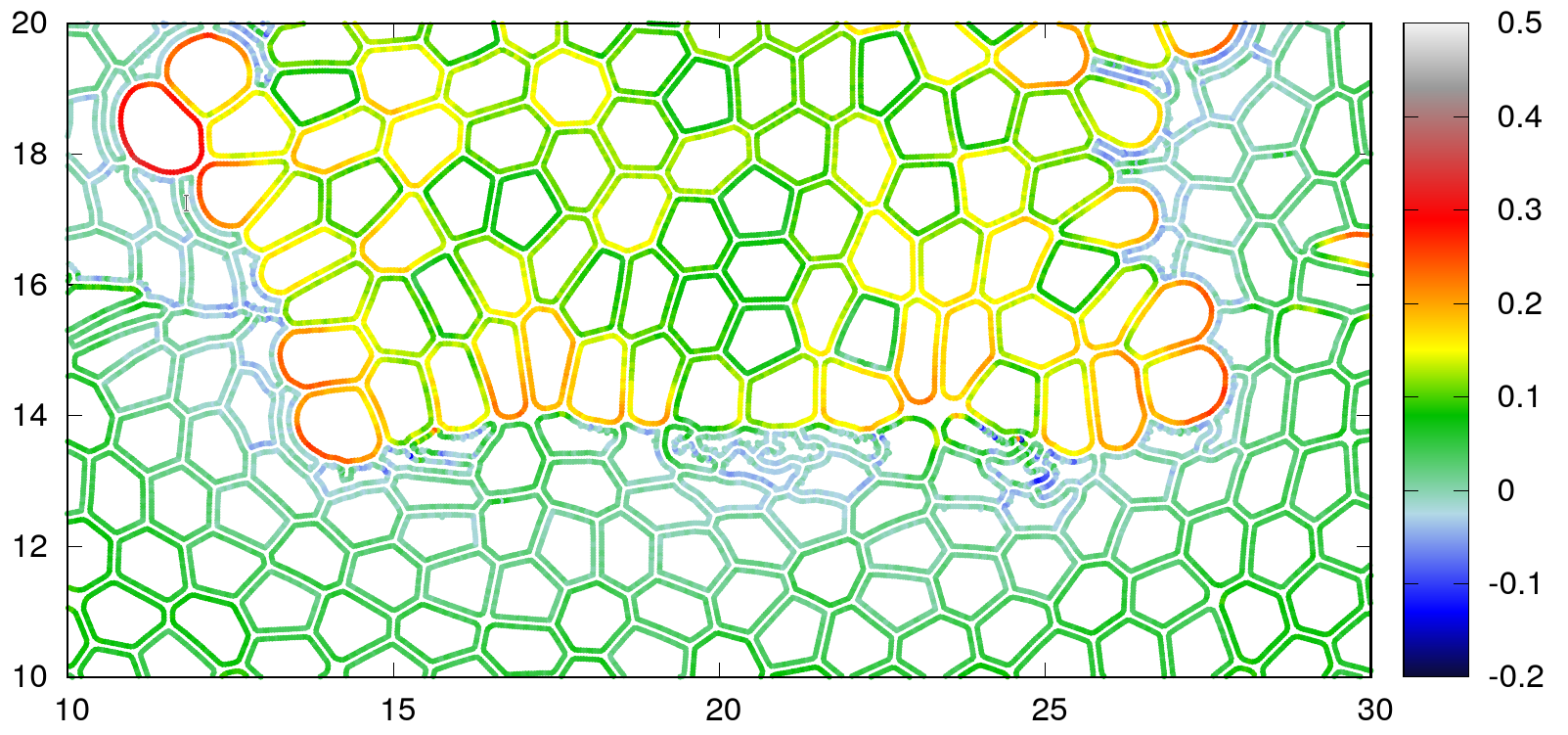}
  \caption{Close-up showing (roughly) the circled area in Fig.~\ref{fig:snap_area_force}a. The color map shows the strain, that is,  relative change in local membrane length as indicated by the color bar (1.0 equals to 100\% strain).
Soft cells orient themselves in the direction of the largest effective stiffness (the soft-stiff boundary). Since strain is given as stress/Young's modulus, it is evident that the soft cells at the boundary layer manifest force dipoles as predicted by Bischofs and Schwarz~\cite{Bischofs2003}.
  }
  \label{fig:closeup}
\end{figure}

Figure~\ref{fig:sim_hl_mu} shows a similar simulation setup as before,
except with two different values of friction $\mu$. Figure~\ref{fig:sim_hl_mu}a
shows the initial conditions, and Figs.~\ref{fig:sim_hl_mu}b and c,
the final states for $\mu\!=\!0.0$~$\mu$g/s and $\mu\!=\!200.0$~$\mu$g/s, respectively.
The tissues in \fig{fig:sim_hl_mu}c grow with open
boundaries.
In both cases, the simulations were run for a time corresponding to 10 division
cycles. As expected, at low inter-membrane friction, there are more cells at
the end of the simulation indicating faster growth. The high inter-membrane
friction system is more porous with slower growth.
Physically, the friction-less system (\fig{fig:sim_hl_mu}b) corresponds to very
early stages of development when junctions have not yet developed and
the latter (\fig{fig:sim_hl_mu}c) to a case when cell
adhesion molecules are present.

To investigate further, we analyzed cellular areas (\fig{fig:area_force_hist}a)
and the forces acting on the cells (\fig{fig:area_force_hist}b) at confluence.
Both distributions display lower total number of cells in the case of the
high friction tissue. The peak in area distributions is just below 100\,$\mu$m$^2$,
which is due to the threshold division area ($A^\mathrm{div}\! =\! 100$ $\mu$m$^2$).
Some of the cell areas have grown past this limit as division occurs only
at discrete time intervals.

The area distribution for $\mu\!=\!0.0$\,$\mu$g/s (\fig{fig:area_force_hist}a) shows
a small peak at $A \! \approx \! 20$\,$\mu$m$^2$ due to the higher number of
collapsed cells in the low friction system. The large-area peak represents
the soft-cell majority with approximately Gaussian shape. This is consistent
with the results from simulations of non-dividing soft colloids~\cite{Astrom2006}. For
$\mu \! = \!200.0$\,$\mu$g/s the distribution develops only a single peak.

The distribution of contact forces can perhaps be best understood by picturing two different phases of tissue:  1) A granular phase in which cell density and tissue structure resemble a granular material near jamming. Then the contact forces have the characteristics of granular force-chains with an exponential force distribution at the high-force. 2) A densely packed phase in which cells have overcome jamming, cell density approaches space-filling, forces have been equilibrated and approach a Gaussian. These aspects are discussed in detail below.

All contact forces would relax to toward zero if proliferation
stopped and forces were measured after a long enough time.
During growth and at low cell-cell friction, the force distribution in
\fig{fig:area_force_hist}b is fairly close to a Gaussian (without negative values).
There is however a hint of two separate peaks;
when friction is increased, the distribution deviates more from
a Gaussian, and its overall structure approaches an exponential
function indicating that at zero friction the tissue is closer to
the dense-packed phase, while at higher friction the tissue
is closer to the granular phase. At high friction, the secondary
peak in force distribution develops more clearly.
Approximately exponential tails
have also been reported in the cell cycle experiments of
Trepat \textit{et al.}~\cite{Trepat2009}. We will return to this
issue in Discussion.

Cells at and above the peaks at larger forces arise as a result
of remaining mechanical frustrations in particular at stiff/soft cell boundaries.
These frustrations are, as expected, more pronounced at high friction,
and more relaxed at low friction. Using tissue with only one type of cells present,
the high-force tails of the distributions vanish entirely as there are no such boundaries.

As the final case,  we study a cluster of soft cells
surrounded by stiff cells to model tumor growth in a healthy tissue
and with closed boundary conditions to be able to study densely packed tissues.

Figures~\ref{fig:snap_area_force}a,b show
morphologies for $\mu\!=\!0.0$\,$\mu$g/s and  at $\mu\!=\!200.0$\,$\mu$g/ at confluence.
\Fig{fig:snap_area_force} shows that the softer (blue) cells
introduced into matrices of stiffer cells grow faster when
inter-membrane friction is low; weaker cell-cell interactions provide
conditions for easier growth. This suggests that inter-cellular
interactions can be an indicator of how well epithelial tissue can
damp the growth of rogue cells that have a higher growth potential.

The high friction system contains a few pores that have formed due to jamming.
Cell sizes are roughly equal within the tumor and inside the matrix but
along the tumor boundary, however, the matrix cells are compressed and the
tumor cells are enlarged.
Further analysis also showed that contact forces are lowest far from the tumour, slightly elevated inside, while largest forces are scattered along the tumor boundary. The matrix cells at the boundary are compressed (i.e. they have a negative membrane strain), while the tumour cells are elongtaed (i.e. they have a positive mebranes strain) near the boundary, Fig.~\ref{fig:closeup}.
This effect is more pronounced in the low friction case.

\begin{figure}[h!t]
	\centering
	\begin{minipage}{\columnwidth}
          \includegraphics[width=0.96\columnwidth]{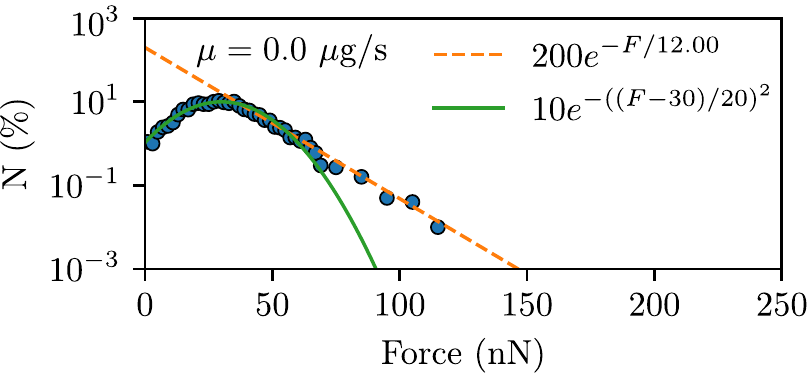}
           \put(-190,37){\colorbox{white!100}{\textbf{(a)}}}
        \end{minipage}\\
        \begin{minipage}{\columnwidth}
          \includegraphics[width=0.96\columnwidth]{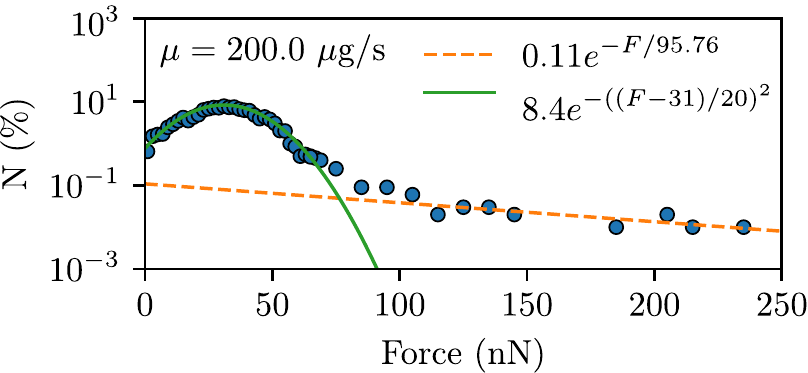}
          \put(-190,37){\colorbox{white!100}{\textbf{(b)}}}
        \end{minipage}
	\caption{Distributions of
          cell-cell contact
          forces ($F$) when a \textit{Soft} cell is initially introduced into a
          tissue of \textit{Stiff}
          cells. (a) $\mu\!=\!0.0$~$\mu$g/s
          and (b) $\mu\!=\!200.0$~$\mu$g/s
          \fig{fig:snap_area_force}b.
          Configurations are shown in Figs.~\ref{fig:snap_area_force}a and
          \ref{fig:snap_area_force}b, respectively.
         }
	\label{fig:mixed_force_distrib}
\end{figure}

To quantify the above, \figs{fig:mixed_force_distrib}a,b show the contact
force distributions. Independent of friction, the distributions are Gaussian
at small forces with an exponential tail. The Gaussian part suggests that
the force distribution is uniform.
The exponential tail, on the other hand,
is a signature of disorder and jamming induced by mechanical frustrations. Here,
it results from large differences in the cell-cell contact forces
along and near the tumor boundary.

Distributions have also been measured for soft colloidal systems
under compression. It is well established that the distribution has an exponential
tail in the vicinity of the jamming transition, see
e.g.,~\cite{Erikson:02eb,Astrom2006,Jose2016} and reference therein.
Experiments by
Jose \textit{et al.} for 3-dimensional packings of soft colloids also show that
the distribution well above the jamming transition
becomes Gaussian~\cite{Jose2016}.
As Fig.~\ref{fig:mixed_force_distrib} shows, the exponential tail is present
at both zero and high friction. The fact that the cells
grow also means that their volumes are not conserved (in contrast to
typical colloids). This is also the case for the cells that are
being pushed and compressed by their neighbors, see
the snapshots in Figs.~\ref{fig:sim_hl_mu} and \ref{fig:snap_area_force}.
What is clearly different here is the distribution at
low forces: The exponential is preceded by a Gaussian distribution. Gaussian
peak has been observed in simulations of soft colloids in two dimensions with zero friction~\cite{Astrom2006}.

In contrast, in the three dimensional experiments of Jose \textit{et al.} the low force part of
the distribution remained almost flat except well above jamming transition.
In addition,
van Eerd \textit{et al.} have reported faster than exponential decay from their high accuracy Monte Carlo simulations~\cite{Eerd2007} although the deviations can be very hard to detect without high accuracy sampling methods.

\section{Discussion}

Using proliferating Madin-Darby canine kidney cells,
Trepat~\textit{et al.}~\cite{Trepat2009}
studied collective migration of cells and came to the conclusion that
long-range traction forces drive collective migration. They measured
traction distributions and found that in all cases the distributions had
exponential tails. Here, we measured the cell-cell contact forces (Figs.~\ref{fig:mixed_force_distrib}) and also found exponential tails.
Similarly to Trepat~\textit{et al.}~\cite{Trepat2009}, the peak of the
distribution appears to be Gaussian. This indicates that our model is
capable of capturing the behavior of real systems, and it confirms that
mechanical aspects can be modeled using a such a coarse-grained approach.

Fritsch \textit{et al.}~\cite{Fritsch2010} point out the conundrum that although
cancers cells are softer than healthy ones, tumors appear as hard lumps.
In particular, Fritsch \textit{et al.}~\cite{Fritsch2010} write:
"At first sight, cell softening is contradictory to the observation that tumours
are rigid masses -- a notion borne out by the fact that breast tumours are often
felt as lumps. Moreover, this apparent softness of tumours would hinder their invasiveness."
As Fig.~\ref{fig:force_distrib} and \ref{fig:snap_area_force} show, our model is capable of demonstrating the effect of invasive soft tissue and when growth starts within a matrix, it will become balanced by the pressure from the surrounding tissues (Fig.~\ref{fig:snap_area_force}), that is, homeostatic pressure.

In separate studies, Basan \textit{et al.}~\cite{Basan2009} and 
Podewitz \textit{et al.}~\cite{Podewitz2015} investigated the role of homeostatic pressure using experiments, numerical models and mesoscale simulations in which cells are represented by two point-particles with potentials for growth, adhesion and exclusion. Constant temperature and pressure ensemble were used for the molecular dynamics simulations~\cite{Podewitz2015}. Due to the representation of cells with central potentials, the approach is not capable of modeling cell-shape changes to any significant degree. As their main conclusion based on their  model and experiments, they showed that negative homeostatic pressure (due to compression in the bulk) is both possible and stable.
They also showed that homeostatic pressure increases linearly with increasing cell compressibility.
Our results agree with this and also demonstrate that the situation is complex.
For example,  there is destruction of healthy cells at the boundary as a result
of concentrated pressure from the cancer cells.
These results indicate why cancer cells may be softer.

Perhaps the most interesting comparison is to the theoretical work of Bischofs and
Schwarz~\cite{Bischofs2003}. In particular, they demonstrated that cells prefer to
orient themselves in the direction of the largest effective stiffness.
This orientational preference
is illustrated in Figures~\ref{fig:force_distrib}b and \ref{fig:snap_area_force}a
by blue ovals (solid line). The softer blue cells are facing a stiff environment
and as a result, they elongate and organize
with their long axis perpendicular to the interface.
Figure~\ref{fig:closeup} shows the local strain which clearly manifests that
force dipoles are present at the boundary.
This is exactly what is predicted by the theory of
Bischofs and Schwarz for clamped (stiff) boundaries~\cite{Bischofs2003}. In the interior,
the cell shapes are more isotropic as is also the case with a free boundary, see
the oval with the dotted
blue line in Figure~\ref{fig:force_distrib}b; the outermost cells are not
in contact with the boundary
yet and hence do not feel it. This is also clearly visible in the simulations
with open boundaries,
Figure~\ref{fig:force_distrib}. In addition, Figures~\ref{fig:force_distrib}b and
\ref{fig:snap_area_force}a show that when going away from a boundary where the softer cells
have elongated, this order may persist for a few cell lines before disappearing depending on the local environment. Thus, forces are generated
away from the immediate boundary. This is in excellent agreement with the experiments of
Trepat \textit{et al.}~\cite{Trepat2009} who argued that force generation is a collective
long-range effect rather than requiring 'leader cells'. The figure also shows that when the interface
between the hard and soft cells has a more complex shape than a straight line, the cellular
shapes and their orientations become more complex. This is also visible in the case of
high cell-cell friction.

As already discussed in Introduction, different vertex, Potts and Voronoi-type models have been used to model cell division. Typically, the focus has been on glass-like behavior in (dense) tissues and topological transitions such as the neighbor exchange T1 transition~\cite{Fletcher2014,Sussman2018} and in some cases such models have been combined even with dynamics~\cite{Sussman2018,Czajkowski2019}. Since the current model gives spontaneously rise to polygonal distributions (as already reported earlier~\cite{Mkrtchyan2014}), it would be possible to use it to study topological changes. That is, however, beyond the scope of the current paper.

Using a vertex model, Bi \text{et al.}~\cite{Bi2015} demonstrated a new rigidity transition at constant density (confluence). As the key parameter, the model has a (cellular) perimeter to area ratio which determines deformability of cells. This parameter has a critical value distinguishing between rigid and fluid-like tissues.
The same effect can be observed in our model. Cells with high internal pressure form a rigid tissue, while deflated cells easily change their perimeter area allowing them to deform and flow past each other. Cell friction also plays a signigicant role in such behaviour.   
Using cellular area/rigidity ratio is not an optimal control parameter here, but
as Fig.~\ref{fig:mixed_force_distrib} shows, the amount of rigidity is governed by cell-cell friction.

\section{Conclusions}

We use the \textit{EpiCell2D} model to study systems of cells of two different
types in 2D. Cell populations are differentiated by their
membrane/cortex stiffness. We showed that this simple difference is
enough, provided internal pressure is identical for both, to favor
soft cell growth. Even if a few soft cells are surrounded by stiff cells,
it is enough for the softer cell to grow rapidly. This effect can be
mitigated by a higher interaction strength between cells.
Force distributions show similarities to non-proliferating
soft colloidal systems. Although not studied in detail here, the model allows for tuning the cell-cell friction, an issue that has recently been raised by Vinuth and Sastry for shear jamming~\cite{Vinutha2016}.

  From the modeling perspective, the \textit{EpiCell2D} approach appears to be
  very versatile. As discussed in Introduction and elsewhere in the text, there
  are several different models around each with their own strengths. \textit{EpiCell2D}
  is, however, able to capture a very wide range of phenomena without additional
  modifications. As we have already shown
  earlier~\cite{Mkrtchyan2014}, it can reproduce the polygonal cell distributions and mitotic
  indices observed experimentally in epithelial systems, and as shown here, force distributions,
  and cellular response to different boundaries agree well with experiments and theory.



\begin{acknowledgments}
MK thanks the Discovery and Canada Research Chairs Programs
of the Natural Sciences and Engineering Research Council of Canada (NSERC)
for financial support.
\end{acknowledgments}


\end{document}